\documentstyle[psfig,epsfig,12pt]{article}
\begin{document}
\baselineskip 0.25in

\begin{center}
\large {\bf The Question of Relaxation in the Wealth Exchange Models} 
\end{center}
\smallskip
\begin{center}
Abhijit Kar Gupta
\end{center}

\begin{center}
{\it Physics Department, Panskura Banamali College\\
Panskura R.S., East Midnapore, WB, India, Pin-721 152}\\
{\em e-mail:}~{kg.abhi@gmail.com}
\end{center}

\abstract{We look at the meaning of 'relaxation' in the wealth exchange models that are 
recently proposed in Econophysics to interpret the wealth distributions. To quantify and 
characterise the process of relaxation, we define an appropriate quantity and evaluate 
that numerically for the systems of many agents. Also, it has been supported heuristically 
by constructing a simple differential equation.} 

\bigskip

\noindent{\bf Introducing the Models}
\bigskip

The wealth exchange models are many agent models for wealth distributions where a randomly 
chosen pair of agents interact and exchange wealth between them in a certain fashion. 
We here deal with the models that are primarily based on the kinetic theory of gas in 
statistical physics. The interactions among agents can be thought of as an elastic collision 
as the total wealth of the interacting agents is kept constant and this in turn 
ensures the total wealth of all the agents to remain conserved. 
There has been a great amount of study \cite{eco} in this kind of wealth exchange models in econophysics 
in recent times \cite{eco}. 

\smallskip
The time evolution of wealth ($w$) in the wealth exchange models can be summarised as follows:

\begin{equation}
w_i(t+1)=w_i(t)+\Delta w
\end{equation}
\begin{equation}
w_j(t+1)=w_j(t)-\Delta w.
\end{equation}

The above is a zero sum process where the wealth of an agent evolves through such a simple rule 
and a distribution of wealth emerges after a certain 'time' $t$. 
It is possible to obtain a wide variety of distributions within this framework namely, 
the exponential Boltzmann-Gibbs type distributions, the gamma type distributions and 
more interestingly the power laws. Power laws that are obtained from fitting the tail ends of the 
wealth distributions of real data are known as {\it Pareto's law} in the economic 
literature for over a hundred years and this is known to possess some universality.

The emergence of a distribution depends on the expression of exchange amount $\Delta w$ which 
stems from the description of the model.
The models developed out of the above principle have been quite useful in 
understanding the wealth distributions of individuals (or that of companies or societies) in 
an economy (see the recent reviews \cite{abhi-rev, yako-rev}). 

In the following, we write down the expressions for the exchange amount $\Delta w$ for 
different models: 

\begin{eqnarray}
\Delta w  & = &\bar\epsilon w_i - \epsilon w_j \label{gambling}\\
&=&(1-\lambda)(\bar\epsilon w_i - \epsilon w_j) \label{fixlam}\\ 
&=&(1-\lambda_i)\bar\epsilon w_i - (1-\lambda_j)\epsilon w_j \label{randomlam}.
\end{eqnarray}

\noindent The first step in the above [eqn.(\ref{gambling})] is for the pure gambling 
model \cite{puregamb} where we write 
the evolution of the wealth of $i$-th agent is given by $w_i(t+1)=\epsilon(w_i(t)+w_j(t))$,
$\epsilon$ being a random number between 0 and 1 drawn from a random number generator.
Note that here, $\bar\epsilon = 1-\epsilon$. 
The exponential Boltzmann-Gibbs type distribution results from this model.
In the next model [in eqn.(\ref{fixlam})], 
the agents have a fixed saving propensity \cite{constlam} introduced through a 
parameter $\lambda$ where the evolution of the $i$-th agent is given by 
$w_i(t+1)=\lambda w_i(t)+\epsilon(1-\lambda)[w_i(t)+w_j(t)]$.
In this, each agent saves $\lambda$-fraction of his/ her wealth and puts the rest for gambling.
It has been shown that the distributions found from this model are of the gamma-type. 
The last step [eqn.(\ref{randomlam})] corresponds to the model where the saving propensity is 
characteristic of an agent {\em i.e.}, the parameter $\lambda$ has been assigned a 
distribution (in our case a uniform distribution in $\lambda$). 
Power laws in wealth distributions are obtained from this random or distributed saving 
model \cite{varlam} and hence this model does attract enhanced attention.

As the system of many agents evolves in time (as suitably defined), it relaxes towards a 
steady state equilibrium so that the distribution of wealth assumes a definite shape.
In the pure 
gambling model and in the model with fixed saving propensity, the idea of relaxation can hardly  
be of interest as the agents do not possess any characteristic feature; they are having equal 
opportunities (either no saving or equal saving) through random interactions. 
We shall, therefore, concentrate on the random saving model out of the three 
as the meaning of relaxation can be rightfully associated with this. 
The relaxation study has been made earlier \cite{prevrelax} on this model but the concept 
and purpose of that had been different from the present study.

\bigskip

\noindent{\bf To Observe Relaxation}
\bigskip

The question is how we may characterise the 'relaxation' of an evolving system. 
If a system is allowed to go towards a steady or fixed state, it will relax (usually 
exponentially) after the external disturbance 
is withdrawn (as is well known in the context of a model spin system in statistical 
physics). As the system approaches
a steady state, the successive values (in time) of a measurable quantity for the whole 
system are supposed to be nearing each other. 
From this idea we check the concept of relaxation in the wealth exchange models in the 
following way.  

The following quantity may be defined as a measure of relaxation: 

\begin{equation}
X(t) = {1\over N}\sum_i|{w_i(t)-w_i(t-1)}|.
\end{equation}

\noindent The quantity $X(t)$ is averaged over $N$ interactions. 
Further, it is  averaged over a number of initial configurations. When the configuration 
averaged value of $X(t)$ is plotted against time $t$ we expect
a graph decaying with time for a system which is supposed to relax to equilibrium.
Here we define one 'time step' to be equal to $N$ interactions on average, where $N$ is 
the number of agents in the system: $t=T/N$, $T$ being the number of $N$ interactions. 

\bigskip

\noindent{\bf Numerical Results}
\bigskip

From the numerical study of the above models it is seen that the system of many agents 
relaxes towards a steady state equilibrium and the wealth of each individual attains a specific 
time independent distribution. 
It will be of interest to see how a system of 
many agents, as a whole, relaxes as it evolves from arbitrary initial distributions.  
The system obviously does not head towards a fixed stationary state. Rather it fluctuates 
around some average value (in equilibrium) in all cases as the exchange of wealth is allowed 
to happen all the time.
It may be emphasised here that the steady state equilibrium states are not affected by the 
choice of initial configurations.
Also, the initial configurations do not have any bearing 
on the overall process of relaxation ({\em i.e.} the shape of the decay curves). 
We check all these numerically (the data are not presented here).

\begin{figure}[h]
\centerline{\psfig{figure=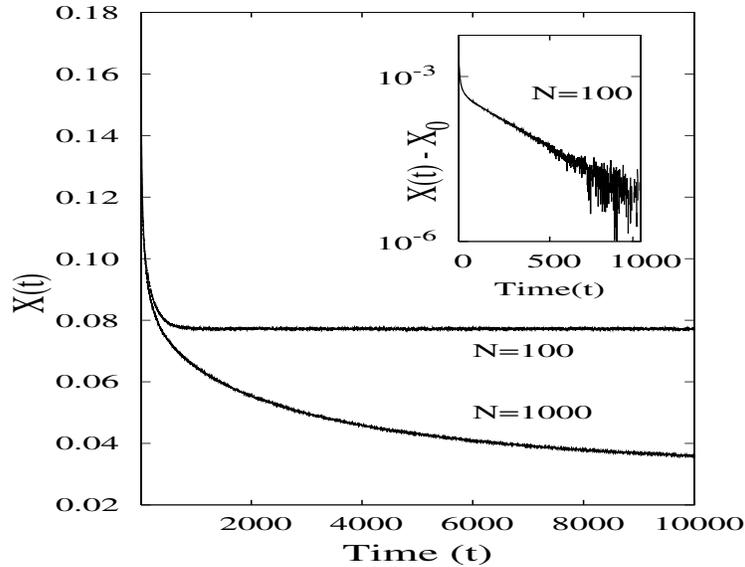, width=4in, height=3in}}
\caption{Shown here is the relaxation in the model with random saving. Initial part is clearly exponential when we subtract the mean value at equilibrium ($X_0$) from 
$X(t)$ and plot in the semi-log scale to demonstrate in the inset for $N$=100.}
\end{figure}

In the numerical simulation, we take $N$=100 in most of the cases as this is sufficient 
for our purpose; only in some cases we take $N=1000$ to demonstrate the size dependence. 
The averaging, in all cases, have been done over $10^4$ initial configurations.
The appearance of exponential decay in relaxation in the 
random saving model \cite{varlam} may be anticipated due to a probable reason that 
the agents possess a characteristic feature (random distribution in the saving 
parameter $\lambda$ in our case). The system is expected to be driven towards a steady state 
equilibrium through a possible exponential decay in such a case. 
In fig.1, we plot averaged value of $X(t)$ against time $t$ where the relaxation seems to be 
of the form: $X(t)=X_0-A.\exp(-t/{\tau})$. Thus to demonstrate the exponential character we 
determine $X_0$ and then subtract that from $X(t)$ to plot in the semi-log scale as shown in
the inset of fig.1.
The relaxation time $\tau$ can be 
identified as the inverse of the slope of the straight line (inset in fig.1).
The relaxation is seen to be dependent upon the system size (number of agents $N$) which can be 
intuitively understood. How it exactly depends on the system size, can be studied later.

\begin{figure}[h]
\centerline{\psfig{figure=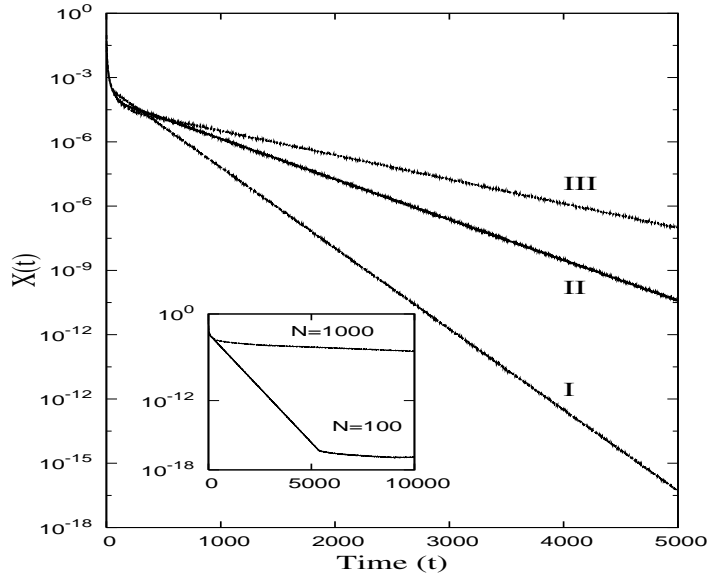, width=4in, height=3in}}
\caption{The quantity $X(t)$ (as defined in the text) is plotted against time $t$ in 
semi-log scale. Demonstration of exponential relaxation in the model with random 
saving $\lambda$ for three cases with $\epsilon = {1\over 2}$:
(I) $0 < \lambda < 1$, (II) $0.5 < \lambda < 1$, (III) $0.7 < \lambda < 1$. 
In the inset, the curve (I) is plotted for a larger time and this shows a saturation regime 
to appear after the pure exponential decay. Another curve in the inset is for a bigger 
system size and for the same parameters.}
\end{figure}

The random saving model has two parameters namely, $\lambda$ and $\epsilon$ where both are taken from a uniform distribution between 0 and 1. It is 
checked through numerical simulation that as the parameter $\lambda$ is taken to be
random (as per the requirement of the model), the other parameter $\epsilon$ may be 
held constant and for that the ultimate wealth distribution does not change. Therefore,
we check the relaxation in this model with different fixed values of $\epsilon$ and out 
of those we find that the relaxation is found to be purely exponential (up to a certain
time) for $\epsilon = {1\over 2}$. 
In fig.2, we demonstrate this and plot the graphs for different widths in the 
random distribution in $\lambda$. Note that the plots are made here without the subtraction of
the saturation value $X_0$ (unlike that in fig.1). The subtraction is not required as the straight 
line portion in the semi-log plot shows a pure exponential decay of the 
type: $X(t)=A.\exp(-t/\tau)$. The system stabilises only after a spell of pure exponential decay.
The relaxation time $\tau$ (inverse of the slope of the straight line) seems to depend on the width 
of distribution in $\lambda$ in a definite way. As the mean in the distribution in $\lambda$ 
increases, the relaxation time increases (slope decreases) which is evident from the graphs. 
We have not investigated here 
how $\tau$ may depend on the width or the mean of the distribution in $\lambda$.
This can be interesting and it will be studied later.

In passing, a similarity of this kind of 
computer model with the well known {\it random resistor network} (rrn) model \cite{kirk} may be 
noted. 
The potential of a node in a resistor network can correspond to the wealth
of an agent. In rrn, the voltage is updated by Kirchhoff's law: 
$V_o^{\prime}=V_o+\Delta V$, 
where $\Delta V=\lambda\sum(V_i-V_o)g_i$, 
$\lambda=\sum g_i$, $g_i's$ being the conductances of the connecting 
resistors. The above updating rule is very similar to that in the wealth exchange models. Note that $\Delta V$ can be positive or negative. 
The quantity that remains conserved here is the 
total current through the connecting resistors in and out of a node. 
Interestingly, here too the relaxation is observed to be exponential. 
To check this, we do a simulation on a random resistor network over a $100\times 100$ square lattice. 
We calculate a similar quantity $X(t)$ as that is done before. 
Here we define

\begin{equation}
X(t) = {1\over N}\sum_{i,j}|{V_{i,j}(t)-V_{i,j}(t-1)}|, 
\end{equation}

\noindent where $V_{i,j}$ is the potential at a node ($i$,$j$).
The resistors are assigned conductance values (inverse of resistance) taken from a uniformly 
random distribution. 

\begin{figure}[h]
\centerline{\psfig{figure=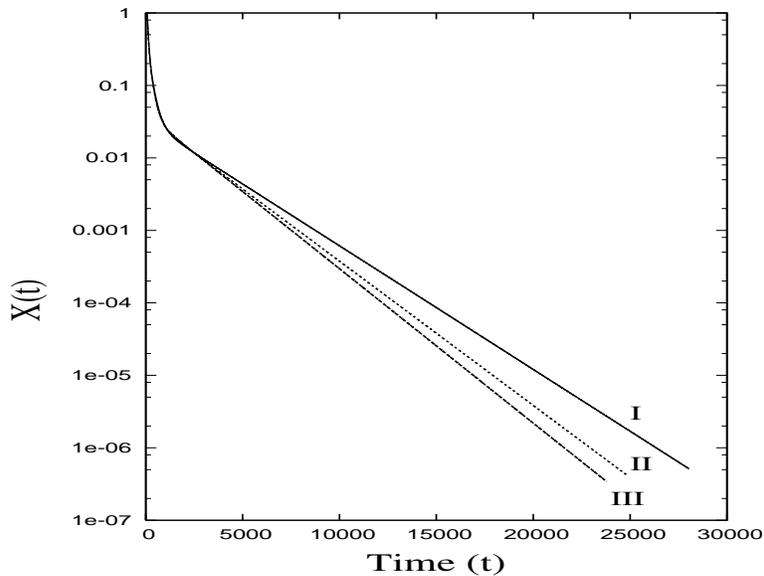, width=4in, height=3in}}
\caption{The relaxation in the random resistor network (rrn) is shown. Exponential decay with 
different relaxation time can be seen from the semi-log plot for different ranges of 
uniform distribution in the resistance/ conductance values: conductances are in the range of (I)
0 to 1, (II) 0.2 to 1 and (III) 0.5 to 1.}
\end{figure}

It is also seen that the relaxation time is dependent upon the width of 
randomness in the distribution of
resistances/ conductances in rrn, quite a similar thing that happens in the wealth distribution model 
with random saving (the results of which is presented in fig.2).
Therefore, a similarity (between the models in two areas) may be drawn in many ways but we
should emphasise here that this is only tentative.
However, the random resistor network model is obviously done on a regular lattice 
whereas the wealth exchange models are usually not done on any lattice. But it is also 
can be checked numerically that in the wealth exchange models, the wealth distribution 
does not change if the agents are taken on a regular lattice. 
Numerical investigations nevertheless suggests that the relaxation time in the 
present wealth exchange model changes due to the presence of a lattice. 

\begin{figure}[htb]
\centerline{\psfig{figure=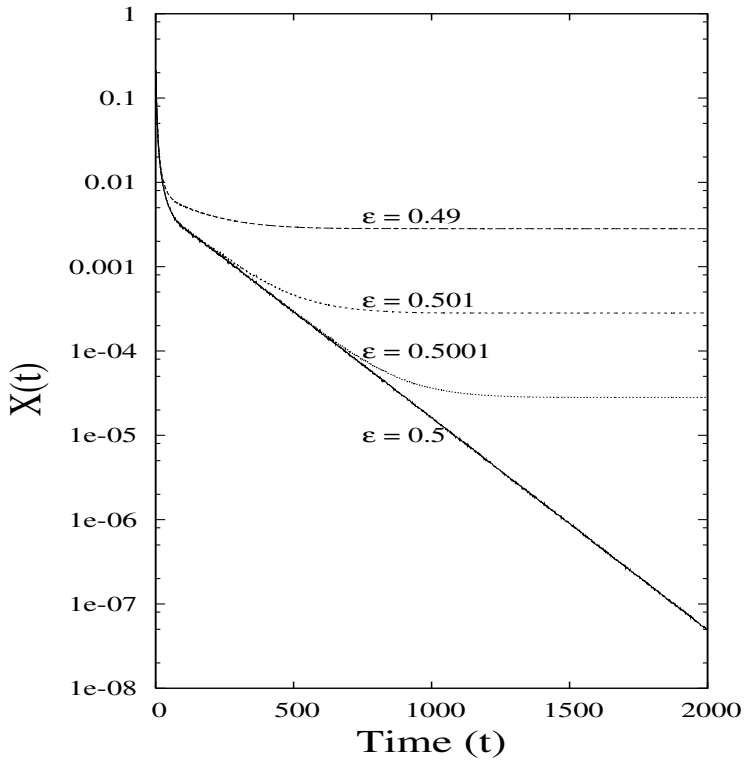, width=4in, height=4in}}
\caption{The relaxation is shown in the random saving model for a set
of values of the parameter $\epsilon$ close to 0.5. The appearance of clear-cut
exponential decay is there for $\epsilon=0.5$ only.}
\end{figure}

We observe that the exponential decay can be prominently demonstrated for the case of
$\epsilon = {1\over 2}$. In this case the system is seen to settle to the lowest possible 
equilibrium value for $X(t)$. So this is a special case. 
We also examined the cases with $\epsilon$ very close to ${1\over 2}$ 
(but not exactly equal to) which are demonstrated in fig.3. 
As the value of $\epsilon$ is taken slightly away from 0.5, the relaxation curve stabilises far 
off from that for $\epsilon={1\over 2}$ (fig.3).

The natural question now is to ask how the wealth of an individual approaches a 
value (for some it grows and for some it decays) on an average with time. This question is 
addressed in the work in \cite{prevrelax}. In the random gambling model and in the 
model with constant $\lambda$ an agent ends up with the same value of wealth on the average. 
However, the model with 
random or characteristic saving \cite{varlam} that we have dealt with here is quite different. 
The agents with higher $\lambda$ (higher saving propensity) ends up with more wealth than 
another with smaller $\lambda$ value. 
This can be understood from common knowledge. The agents with higher saving tendencies ends up 
with accumulating more wealth than those with smaller saving tendencies.
As we start from any arbitrary initial
configuration, the wealths of some agents thus grow towards some higher values and that 
for others decay towards some lower values. This decay or growth is 
exponential (as this is also checked numerically). The growth or decay of all the agents 
are reflected in the relaxation of the entire system. A heuristic argument that follows, may
may be helpful in understanding the general nature of relaxation and that for the special 
case of $\epsilon={1\over 2}$. 

\bigskip
\noindent{\bf Heuristic Arguments on the Nature of Relaxation}
\bigskip

Let us consider a general model which can 
correspond to any of the models stated earlier as we tune the parameters in it appropriately.
A discussion over this based on the transfer matrix approach is given in \cite{abhi-gen}.
The algorithm of wealth exchange in the general model is the following: 

\begin{eqnarray}
w_i(t+1) & = & \epsilon_1 w_i(t)+\epsilon_2 w_j(t) \label{algo}\\
w_j(t+1) & = & (1-\epsilon_1)w_i(t)+(1-\epsilon_2)w_j(t) \label{algo1},
\end{eqnarray}

\noindent where the parameters $\epsilon_1$ and $\epsilon_2$ can be positive or negative 
and they can be related to the actual parameters in the models under consideration.
Note that the values of the above parameters are supposed to be uniformly distributed in 
some range. For example, if $\epsilon_1$ and $\epsilon_2$ both are uniform random numbers 
between 0 and 1, a little analysis suggests that $w_i$ in eqn.(\ref{algo}) will be distributed
uniformly in the range between $-w_i$ and $(w_i+w_j)$.
Thus $w_i$'s can be thought of continuous variables in a certain range. The $w_i$'s, however,
can take only positive values in the models by design. Therefore, we attempt to construct 
differential equations out of the above mentioned coupled 
equations [eqn.(\ref{algo}) \& (\ref{algo1})] assuming that the change in the wealth, 
$\Delta w_i = w_i(t+1)-w_i(t)$ occurs in time $\Delta t$=1. 
The time steps $\Delta t$ can be adjusted according to our convenience.

\noindent We arrive at the following differential equation:

\begin{equation}
{d^2w_i\over dt^2} + (1+\epsilon_2-\epsilon_1){dw_i\over dt} = 0.
\label{diffeqn}
\end{equation}

\noindent The solution of the above homogeneous differential equation is of the following form: 
\begin{equation}
w_i(t) = a + b \exp(-k.t),
\label{sol}
\end{equation}
\noindent where $k = (1+\epsilon_2-\epsilon_1)$; $k$ can be positive or negative depending 
on the choice of the parameters, $\epsilon_1$ and $\epsilon_2$. Therefore, the above
solution can be seen to be either growing or decaying exponentially from or to a certain
value. 

\noindent Now we think of the wealth exchange model with random saving as a special case. 
The wealth of the $i$-th agent evolves through the following way:
\begin{equation}
w_i(t+1)=\lambda_i w_i(t)+\epsilon[(1-\lambda_i)w_i(t)+(1-\lambda_j)w_j(t)],
\label{randomsav}
\end{equation}
\noindent where $i$-th and $j$-th agents save $\lambda_i$ and $\lambda_j$ fractions of their
wealths respectively, at time $t$.
In view of the general equation as mentioned above [eqn.(\ref{algo})], 
we have $\epsilon_1=\lambda_i+\epsilon[(1-\lambda_i)$ and 
$\epsilon_2=\epsilon[(1-\lambda_j)$. 
Thus for the random saving model, we obtain the following expression for $k$ for the choice of 
$\epsilon = {1\over 2}$: 

\begin{equation}
k=1+\epsilon_2-\epsilon_1=1-{1\over 2}(\lambda_i+\lambda_j).
\end{equation} 

As we consider the distributions in $\lambda$ such that
$0 < \lambda_i,\lambda_j < 1$, the value of $k$ must always be positive ($k > 0$). 
Hence the solution [eqn.(\ref{sol})] always decays exponentially for this 
special case for any values of $\lambda$'s as long as they are bounded between
0 and 1. Infact, now we may intuitively understand how a pure exponential 
relaxation may appear in the case for $\epsilon = {1\over 2}$ which is shown in fig.2. This fact 
is even more evident from the demonstration in fig.3.

In conclusion, we have tried to establish the nature of relaxation in the 
wealth exchange models of a certain class by studying the relaxation in random saving model. 
The nature of relaxation is found to be exponential (followed by a saturation) and this is possibly 
true for other models (that are not discussed here) based on the similar principles.
The relaxation process (and the relaxation time) can be greatly manipulated by tuning the 
parameters, $\lambda$ and $\epsilon$ that are involved in random saving model.
The relaxation is shown to be purely 
exponential for the special choice of $\epsilon = {1\over 2}$ in this model. The claims are
made from numerical simulation results and are supported by heuristic arguments.

\bigskip
\noindent{\bf Acknowledgement}
\smallskip

The author wants to thank {\em D. Stauffer} and {\em B.K. Chakrabarti} for their valuable 
comments and critical remarks on the work and on the manuscript.

\end{document}